\documentclass[
aps,%
8pt,%
final,%
notitlepage,%
oneside,%
onecolumn,%
nobibnotes,%
nofootinbib,%
superscriptaddress,%
noshowpacs,%
]%
{revtex4}

\begin{document}
\selectlanguage{english}
\noindent {\it ASTRONOMY REPORTS, 2015, Vol. 59, No 9}\bigskip\bigskip  \hrule\smallskip\hrule 
\vspace{15mm}

\title{\uppercase{A Numerical Model for Accretion\\ in Intermediate Polars  
with Dipolar\\ Magnetic Fields}}

\author{\bf \copyright $\:$  2015. \quad\firstname{P.~B.}~\surname{Isakova}}
\email{isakovapb@inasan.ru}
\affiliation{Institute of Astronomy, Russian Academy of Sciences, ul. Pyatnitskaya 48, 
Moscow 119017, Russia}%
\author{\bf   \firstname{A.~G.}~\surname{Zhilkin}}
\affiliation{Institute of Astronomy, Russian Academy of Sciences, ul. Pyatnitskaya 48, 
Moscow 119017, Russia}%
\affiliation{Chelyabinsk State University, Chelyabinsk, Russia}%
\author{\bf  \firstname{D.~V.}~\surname{Bisikalo}}
\affiliation{Institute of Astronomy, Russian Academy of Sciences, ul. Pyatnitskaya 48, 
Moscow 119017, Russia}%

\begin{abstract}

{\footnotesize \qquad\quad Received   April 3, 2015;
$\;$ in final form April 10, 2015}\bigskip\bigskip

\noindent
A three-dimensional numerical model for an accretion process investigation in the magnetosphere of a white dwarf in
magnetic cataclysmic variables is developed. The model assumes that the white dwarf has a dipole magnetic
field with its symmetry axis inclined to the rotation axis. The model is based on the equations of
modified MHD, that describe the mean flow parameters in the wave MHD turbulence. Diffusion of the
magnetic field and radiative heating and cooling are taken into account. The suitability of the model
is confirmed by modeling the accretion in a typical intermediate polar. The computations show that
a magnetosphere forms around the accretor, with the accretion occurring via columns. The accretion
columns have a curtain-like shape, and arc-shaped zones of energy release form on the surface of the white
dwarf in the magnetic poles area as a result of the matter infall.

DOI: 10.1134/S106377291509005X
\end{abstract}

\maketitle

\newpage

\section{INTRODUCTION}

Studies of accretion processes are the
most important and topical problems of modern
astrophysics. In many cases, the intrinsic magnetic
field of the accretor plays a substantial role in this
process. We consider here magnetic cataclysmic
variables as investigation objects. These are close binary systems consisting of a 
low-mass late-type star (donor) and a white dwarf
(accretor) \cite{Warner1995}, the donor fills its Roche lobe.
The pressure gradient at the inner Lagrangian point
L$_1$ is not balanced by gravity, and matter starts to
flow into the Roche lobe of the compact object. Two
main types of magnetic cataclysmic variables can be
distinguished: polars and intermediate polars. In polars the
accretorhas a substantial magnetic field
($> 10^6$ G) and magnetosphere that extends to the
inner Lagrangian point L$_1$ , prevent the formation of
an accretion disk. The accretion proceeds along
magnetic-field lines onto the magnetic poles of the
accretor. In intermediate polars, the magnetic field of
the accretor is relatively weak ($10^4$--$10^6$ G), and an
accretion disk can form in the system, whose inner
radius is limited of the size of the magnetosphere.
The interaction of the disk material and the 
magnetic field of the accretor leads to the formation of a complex flow
structure, that can include accretion columns and
belts. The flow structure in intermediate polars also
depends strongly on the inclination of the magnetic
axis, the magnetic field geometry, and the rotational
velocity of the accretor.

The three-dimensional (3D) numerical model of the
accretion taking into account the magnetic field of
the accretor was for the first time developed in \cite{Koldoba2002}. This model
made it possible to describe in detail the 3D structure
of the flow in the magnetosphere of the gravitating
object, where the magnetic field is dominant. The
following studies of these authors \cite{Romanova2003, Romanova2004a, Romanova2004b} 
presented the results of 3D
numerical modeling of plasma accretion onto a gravitating object 
with a dipole magnetic field whose axis
of symmetry is not aligned with the rotational axis
of the star. For the first time 3D MHD simulation of disk
accretion onto a star with a complex magnetic field
geometry was carried out in \cite{Long2007,Long2008}. More complex
magnetic field configurations were considered in \cite{Romanova2011},
taking into account the octupolar component. These
models were used to study disk accretion onto young
T Tauri stars.

In the series of our papers \cite{zb2009, ZhilkinMM2010, zbASR2010, zbMFG, zbSMF, zbbUFN, 
bycam1} (see also the monograph \cite{mcbook}), we developed a 3D numerical model for
mass transfer in semidetached binary systems taking into 
account the magnetic field of the accretor.
The model assumes that the plasma dynamics are
determined by the slow mean flow, against the background 
of that highspeed MHD waves propagate.
A strong external magnetic field acts as an efficient
fluid and interacts with the plasma. The model
takes into account the inclination of the magnetic
axis relative to the rotational axis, the diffusion of
the magnetic field, and radiative heating and cooling.
This model allowed us for the first time to achieve a
selfconsistent description of the MHD flow structure
in close binary systems, including such characteristic
features as the accretion disk, the magnetosphere of
the accretor, accretion columns, etc. However, over
our limited computer resources, we focused on studies of 
the outer regions of the accretion disk, that
can be observed using classical astronomy methods.
We did not consider the flow structure near the surface of 
the accretor in detail before. In our present paper,
we describe the numerical model that allows detailed
study of the flow characteristics in the vicinity of
the accretor magnetosphere in the frame of our
general approach. This makes it possible, in particular,
to study the penetration of plasma into the
magnetospheres of white dwarfs and neutron stars in
more detail.

The paper is organized in the following way. Section $2$
describes the model, and Section $3$ is the numerical
method. The results of the 3D numerical modeling
are presented in Section $4$. We discuss the main
results of our study in the Conclusion.

\section{DESCRIPTION OF THE MODEL}

The object of our study is a close binary system
with the parameters of the typical intermediate polar 
\cite{Warner1995}. The donor (red dwarf) has a 
mass of $M_d = 0.1~ M_{\odot}$ 
and an effective temperature of $4000~\text{K}$. The
mass of the white dwarf is $M_a = 0.8~M_{\odot}$. The orbital
period of the system is $P_\text{orb} = 1.6~ hr$, and the component 
separation is $A = 0.7~R_{\odot}$. We assume that
the magnetic field of the white dwarf can be correctly described
by a dipole field. The surface field of
the white dwarf is varied from $8$ to $80~\text{kG}$. The
inclination of the magnetic axis to the rotational axis
is $30^\circ$.
To describe the flow structure in our numerical
model, we have used a noninertial reference frame
corotating with the binary system with angular velocity
$\Omega = 2 \pi / P_\text{orb}$ relative to its center of mass. The
field strength in this system is described by the Roche
potential
\begin{equation}\label{eq-Phi} 
 \Phi = 
 -\frac{G M_a}{|{\bf r} - {\bf r}_a|} - 
 \frac{G M_d}{|{\bf r} - {\bf r}_d|} - 
 \frac{1}{2} \left[ 
 {\bf \Omega} \times \left( {\bf r} - {\bf r}_c \right) \right]^2,
\end{equation}
where $G$ is the gravitational constant, ${\bf r}_a$~--- the radius
vector of the center of the accretor, ${\bf r}_d$~--- the radius vector
of the center of the donor, and ${\bf r}_c$~--- the radius vector of
the center of mass of the binary system. We use the Cartesian
coordinate system $(x, y, z)$ with the origin coincident with the center of the accretor. 
The center of the donor is located at $(-A, 0, 0)$,
and the $z$ axis is directed along the axis of the system rotation, ${\bf\Omega} = (0, 0, \Omega)$.

The vector of the dipole magnetic field is given by
\begin{equation}\label{eq-Bs} 
 {\bf B}_{*} = 
 \frac{3 \left( \bm{\mu} \cdot {\bf r} \right) {\bf r}}{r^5} - 
 \frac{\bm{\mu}}{r^3}, 
\end{equation}
where $\bm{\mu}$ is the magnetic moment of the accretor. To
reduce the numerical errors, we represent the total
magnetic field ${\bf B}$ as a superposition of the intrinsic
magnetic field of the accretor ${\bf B}_{*}$ and the field ${\bf b}$ 
induced by currents in the plasma: ${\bf B} = {\bf B}_{*} + {\bf b}$ 
\cite{Tanaka1994}. 

In the general case, the rotation of the accretor is
asynchronous, and it is characterized in the reference
frame selected by the angular velocity ${\bf \Omega}_a$. In the case
of synchronous rotation, ${\bf \Omega}_a = 0$. We consider here
the case when the rotational axis of the accretor is
aligned with the rotation axis of the binary system. 
Thus, the magnetic field of the accretor
is nonstationary:
\begin{equation}\label{eq-Bs2}
 \pdiff{{\bf B}_{*}}{t} = \rotor ({\bf v}_{*} \times {\bf B}_{*}),
\end{equation}
where ${\bf v}_{*} = {\bf \Omega}_a \times ({\bf r} - {\bf r}_a)$ is the velocity of the
magnetic field lines of the accretor.

The plasma in the vicinity of the magnetospheres
of the magnetic white dwarfs in cataclysmic variables
is magnetized \cite{zbbUFN} and moves in the external 
magnetic field. The intrinsic magnetic field of the plasma
${\bf b}$ is then much weaker than the intrinsic magnetic
field of the white dwarf ${\bf B}_{*}$. In this case, the velocity
of the plasma can be much lower than the propagation velocity of 
MHD waves. In regions of strong
magnetic fields or low density, the velocity of propagation of 
Alfv\'en and magnetosonic waves can
be even relativistic. Over the dynamical time scale,
such MHD waves have time to cross the accretion
stream in the longitudinal and transverse directions
many times. Therefore, the plasma dynamics in the
stream can be considered in the framework of modified
magnetogasdynamics as a sort of mean flow against
the background of a wave MHD turbulence. 
To describe the motion of the plasma in this case we define
the rapidly propagating MHD fluctuations
and apply a certain averaging procedure over the
ensemble of wave pulsations. Such model was
developed by us earlier in \cite{zbSMF, zbbUFN}.

Strictly speaking, this model is correct only in
the presence of a strong external magnetic field, as
is the case of polars and in the magnetospheres of
intermediate polars. However, the results of 
calculations \cite{zbuAIP2013} have demonstrated that this model is sufficiently
universal. With an appropriate choice of parameters
(for example, the parameter determining the efficiency
of the wave turbulence), this model also well describes
the flow structure in the case of weak magnetic
fields. Accordingly, we adopted it as the basis for
our description of accretion in magnetic cataclysmic
variables in the vicinity of the white dwarf magnetosphere.

Taking into account the magnetic field, the flow of
matter in a close binary system may be described by
the system of equations\cite{zbbUFN}:
\begin{equation}\label{eq-rho1} 
 \pdiff{\rho}{t} + \nabla \cdot \left( \rho {\bf v} \right) = 0,
\end{equation}
\begin{equation}\label{eq-v1} 
 \pdiff{{\bf v}}{t} + \left( {\bf v} \cdot \nabla \right) {\bf v} =  
 -\frac{\nabla P}{\rho} - 
 \frac{{\bf b} \times \rotor {\bf b}}{4 \pi \rho} -  
 \nabla \Phi +  2 \left( {\bf v} \times {\bf \Omega} \right) - 
 \frac{\left( {\bf v} - {\bf v}_{*} \right)_{\perp}}{t_w},
\end{equation}
\begin{equation}\label{eq-b1} 
 \pdiff{{\bf b}}{t} = 
 \rotor \left[ 
 {\bf v} \times {\bf b} +  
 \left( {\bf v} - {\bf v}_{*} \right) \times {\bf B}_{*} - 
 \eta\, \rotor {\bf b} 
 \right],
\end{equation}
\begin{equation}\label{eq-s1} 
 \rho T \left[  
 \pdiff{s}{t} + 
 \left( {\bf v} \cdot \nabla \right) s 
 \right] =  
 n^2 \left( \Gamma - \Lambda \right) +  
 \frac{\eta}{4 \pi} \left( \rotor {\bf b} \right)^2.
\end{equation}
where $\rho$ is the density, ${\bf v}$~--- the velocity, $P$~--- the pressure,
$s$~--- the entropy per unit mass, $n = \rho/m_p$~--- the number
density, $m_p$~--- the proton mass, $\eta$~--- the coefficient of 
magnetic viscosity, and $\Gamma$ and $\Lambda$~--- the radiative heating and
cooling functions, respectively. The density, entropy
and pressure are related by the equation of state of an
ideal gas,
\begin{equation}\label{eq-s2} 
 s = c_\text{V} \ln(P / \rho^{\gamma}), 
\end{equation}
where $c_\text{V}$ is the specific heat capacity at constant volume
and $\gamma = 5/3$ is the adiabatic index. The last term
in the equation of motion \eqref{eq-v1} describes the force of
the white dwarf magnetic field acting on the plasma,
that influences the plasma velocity perpendicular to
the magnetic field lines \cite{zbSMF, zbbUFN}. The time scale for
the decay of the transverse velocity is
\begin{equation}\label{eq-tw} 
 t_{w} = \frac{4 \pi \rho \eta_{w} }{B^2_{*}},
\end{equation}
where $\eta_{w}$~--- is the coefficient of magnetic viscosity due
to wave MHD turbulence.

The numerical model takes into account the effects 
of diffusion of the magnetic field [in \eqref{eq-b1} and \eqref{eq-s1}]
caused by magnetic reconnection and the dissipation of
currents in turbulent vortices \cite{Bisnovatyi-Kogan76, zb2009, 
zbbUFN}, magnetic
buoyancy \cite{Campbell97, zb2009, zbbUFN} and wave MHD turbulence 
\cite{zbSMF, zbbUFN}. The coefficient of the wave viscosity is given by
\begin{equation}\label{eq-etaw} 
 \eta_{w} = \alpha_{w} \frac{l_{w} B_{*}}{\sqrt{4 \pi \rho}},
\end{equation}
where $l_{w} = B_{*} / |\nabla B_{*}|$ is the characteristic spatial
scale of the wave pulsations, and $\alpha_{w}$~--- a 
dimensionless factor that is close to unity that determines the
efficiency of the wave diffusion. The diffusion of the
magnetic field is nonlinear in whole.

\section{NUMERICAL METHOD}

The system \eqref{eq-rho1}--\eqref{eq-s1} is quite difficult to solve 
numerically directly. Therefore, it is convenient to 
divide it according to physical processes into simpler
subsystems. Each subsystem can be solved using
specific numerical methods. Let us suppose that we
know the distribution of all values in the computational 
domain at time $t^{n}$. To obtain the values at
the next time step, corresponding to the time $t^{n+1} = t^{n} + \Delta t$, 
we apply an algorithm with five sequential
steps, described briefly below.

In the first step, we distinguish the subsystem of
equations describing the dynamics of the plasma in
its own magnetic field:
\begin{equation}\label{eq-method-1}
 \pdiff{\rho}{t} + \nabla \cdot \left( \rho{\bf v} \right) = 0,
\end{equation}
\begin{equation}\label{eq-method-2}
 \pdiff{{\bf v}}{t} + \left({\bf v} \cdot \nabla \right) {\bf v} =
 -\frac{\nabla P}{\rho} -
 \frac{{\bf b} \times \rotor{\bf b}}{4\pi\rho},
\end{equation}
\begin{equation}\label{eq-method-3}
 \pdiff{{\bf b}}{t} = \rotor \left( {\bf v} \times {\bf b} \right),
\end{equation}
\begin{equation}\label{eq-method-4}
 \pdiff{s}{t} + ({\bf v} \cdot \nabla) s = 0.
\end{equation}
The form of this subsystem coincides with the equations 
of ideal magnetogasdynamics. This subsystem 
can be solved numerically using the higher-order
Godunov-type difference scheme described below in
this section.
In the second step, the variations of the gas 
velocity due to external forces (the Coriolis force and
the gradient of the Roche potential) are taken into
account:
\begin{equation}\label{eq-method-5}
 \pdiff{{\bf v}}{t} = 2 ({\bf v} \times {\bf \Omega}) - \nabla\Phi.
\end{equation}
In this step, the remaining variables are taken to be
constant. The Roche potential $\Phi$ is time independent.
Therefore, we compute new values for the velocity ${\bf v}$
using the analytic solution of this equation in the interval
$t^n \le t \le t^{n+1}$.

The third step of the algorithm considers
the deceleration force during the motion of the
plasma across magnetic field lines, as well as the
generation of magnetic field due to this motion. The
corresponding equations can be written
\begin{equation}\label{eq-method-6}
\begin{gathered}
 \pdiff{{\bf v}_{\perp}}{t} = 
 -\frac{({\bf v} - {\bf v}_{*})_{\perp}}{t_w}, \\ 
 \pdiff{{\bf b}}{t} = 
 \rotor \left[ 
 ({\bf v} - {\bf v}_{*})_{\perp} \times {\bf B}_{*} 
 \right].
\end{gathered} 
\end{equation}
We compute the new values of the velocity ${\bf v}$ and
magnetic field ${\bf b}$ using analytical solutions of these
equations in the interval $t^n \le t \le t^{n+1}$.

In the fourth step of the algorithm, we take into
account the effects of the magnetic field diffusion. The
corresponding equation is
\begin{equation}\label{eq-method-7}
 \pdiff{{\bf b}}{t} = -\rotor \left( \eta \rotor {\bf b} \right).
\end{equation}
This equation is nonlinear in our model. Therefore, it
was solved numerically using an implicit, locally one-dimensional
method with a factorisable operator
\cite{Samarsky1989, mcbook}. 

Finally, the fifth step includes the effects
of radiative heating and cooling, as well as heating
due to current dissipation. These processes are 
described by the right hand side of Eq. \eqref{eq-s1}.

Let us describe the method used to solve the 
hyperbolic subsystem \eqref{eq-method-1}--\eqref{eq-method-4} in more detail. We
can rewrite these equations in conservative form in
the Cartesian coordinates $x^1 = x$, $x^2 = y$, $x^3 = z$ as
follows: 
\begin{equation}\label{eq-method-8} 
 \pdiff{\cal U}{t} + 
 \pdiff{{\cal F}_1}{x^1} + 
 \pdiff{{\cal F}_2}{x^2} + 
 \pdiff{{\cal F}_3}{x^3} = 0.
\end{equation}
Here, ${\cal U}$ and ${\cal F}_k$ (where the subscript $k$ runs through
the values 1, 2, 3) denote vectors of conservative
variables and fluxes, defined by the expressions
\begin{equation}\label{eq-method-9} 
 {\cal  U} = 
 \left(
 \begin{matrix}
 \rho \\ 
 \rho{\bf v} \\ 
 {\bf b} \\ 
 \rho s
 \end{matrix}
 \right), \quad
 {\cal F}_k = 
 \left(
 \begin{matrix}
 \rho v_k \\ 
 \rho{\bf v} { v}_k + {\bf n}_k \left( P + {\bf b}^2 / 8\pi \right) - {\bf b} b_k / 4\pi \\ 
 {\bf b} v_k - {\bf v} b_k \\ 
 \rho s v_k
 \end{matrix}
 \right),
\end{equation}
where ${\bf n}_k$ are unit vectors directed along the axes of
the Cartesian coordinate system.

We now transform the variables $x^k$ in these
equations into the new variables $\xi^k$ using the transformation of coordinates. In the new
curvilinear coordinates $\xi^k$, it is convenient to 
introduce the local basis vectors ${\bf e}_k = 
{\partial {\bf r}}/{\partial \xi^k}$, directed tangentially
to the corresponding coordinate lines. In general,
this basis is nonorthogonal and nonnormalized. In
addition to these vectors, we can also consider the
vectors of the reciprocal (dual) basis ${\bf e}^1 = 
{\bf e}_2 \times {\bf e}_3$, 
${\bf e}^2 = {\bf e}_3 \times {\bf e}_1$ and ${\bf e}^3 = 
{\bf e}_1 \times {\bf e}_2$, , that are orthogonal
to the corresponding coordinate lines. The vectors
of the reciprocal basis are also nonorthogonal
and nonnormalized in the general case. The Jacobian
of the coordinate transformation can be written $Q = {\bf e}_1 \cdot 
({\bf e}_2 \times {\bf e}_3)$. 

In the new variables, the system of equations 
\eqref{eq-method-8} acquires the form
\begin{equation}\label{eq-method-10} 
 \pdiff{{\cal U}}{t} + 
 \frac{1}{Q} 
 \pdiff{{\cal H}_1}{\xi^1} + 
 \frac{1}{Q} 
 \pdiff{{\cal H}_2}{\xi^2} + 
 \frac{1}{Q} 
 \pdiff{{\cal H}_3}{\xi^3} = 0,
\end{equation}
where the fluxes
\begin{equation}\label{eq-method-11} 
 {\cal H}_k = \sum\limits_{i=1}^{3} e^k_i {\cal F}_i,
\end{equation}
and $e^k_1$, $e^k_2$ and $e^k_3$ denote the components of the
vectors ${\bf e}^k$  of the reciprocal basis in Cartesian 
coordinates. We obtained a numerical solution of this
system of equations using a high-order, Godunov-
type difference scheme \cite{ZhilkinMM2010}, providing a third-
order approximation in the spatial variable in the area
of the smooth solution and a first order approximation
in time. With appropriate boundary conditions, this
difference scheme ensures accurate satisfaction of the
conservation laws for the physical quantities in the space of the original variables 
($x$, $y$, $z$). The stability of the scheme is provided
by the limited time step $\Delta t$ (the Courant--Friedrichs--
Lewy condition).

In the computations considered below, we use the
spherical coordinates $\xi^1 = r$, $\xi^2 = \theta$, $\xi^3 = \varphi$ as 
curvilinear coordinates, that are related to the Cartesian
coordinates by the expressions
\begin{equation}\label{eq-coord} 
 x = r \sin\theta \cos\varphi, \quad
 y = r \sin\theta \sin\varphi, \quad
 z = r \cos\theta.
\end{equation}
The solution was obtained in the computational domain ($0.015A \le r \le 0.08A$, $0 \le \theta 
\le \pi$, $0 \le \varphi \le 2\pi$).
We use a $N_r \times N_\theta \times N_\varphi = 128 \times 128 \times 256$ grid in
the numerical computations.

We use the following boundary and initial conditions. 
A free inflow condition is specified at the
inner boundary, corresponding to the surface of the
accretor. We adopt the magnetic field condition
${\bf b} = 0$. We neglect additional heating of the matter
due to the absorption of radiation from the accretion
zones. The fields at the surface of the accretor are
taken to be typical for intermediate 
polars. In such systems, mass transfer leads to the
formation of an accretion disk. Therefore, we specify
conditions corresponding to the distributions of 
variables in an accretion disk at the outer boundary of the
computational domain. The vertical density 
distribution is defined using the condition of hydrostatic
equilibrium in an isothermal disk with temperature
$10000~\text{K}$. We apply the conditions $v_\varphi = v_\text{K}$ 
and $v_r = -0.1 v_\text{K}$ for the velocities, where $v_\text{K}$ is the velocity
of the Keplerian rotation. In the equatorial plane of the
disk, we specify the density to be $\rho = 10^{-3} \rho(\textrm L_1)$,
where $\rho(\textrm L_1)$ is the density at the inner Lagrangian
point L$_1$ corresponding to a mass transfer rate 
$\dot{M} = 10^{-10}~M_\odot/\text{yr}$.
The initial conditions in the computational domain 
are following: initial density $\rho = 10^{-6} \rho(\textrm L_1)$, 
initial velocity ${\bf v} = 0$, initial temperature
$T = 10000~\text{K}$, and initial magnetic field ${\bf b} = 0$.

We use the 3D parallel code \cite{zbbUFN, ZhilkinMM2010} for the
numerical simulation. The computations for all the
models were continued until the 
quasistationary regime begins, it is defined by approximate (up to 1\%)
constancy of the total mass in the computational
domain. The computations were carried out at the 
computer cluster of the Joint Supercomputer Center 
of the Russian Academy of Sciences
using 512 processors.

\section{COMPUTATION RESULTS}
Here we present the results of our simulation of
the flow structure for magnetic fields at the surface of
the accretor of $B_a = 8~\text{kG}$ (Model 1) and $B_a = 80~\text{G}$
(Model 2). The inclination of the magnetic axis to
the rotational axis ($z$) was $30^\circ$ in both cases. Since
the radius of the magnetosphere exceeds the radius
of the accretor in both models, the accretion has
a magnetogasdynamical rather than gas dynamical
character.

\begin{figure}[ht!]  
\centering 
\begin{tabular}{c c}
\hbox{\includegraphics[width=0.45\textwidth]{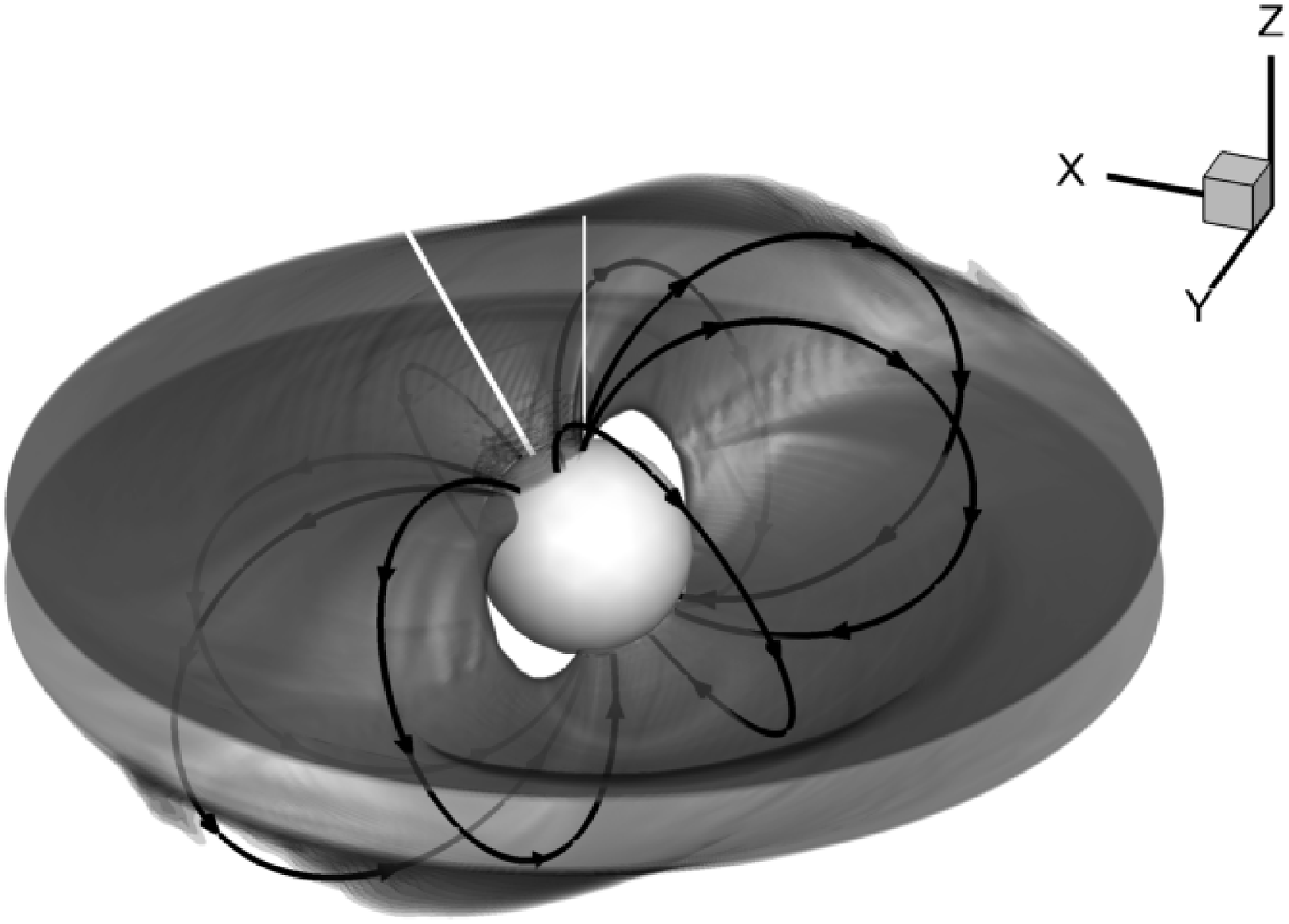}} &
\hbox{\includegraphics[width=0.45\textwidth]{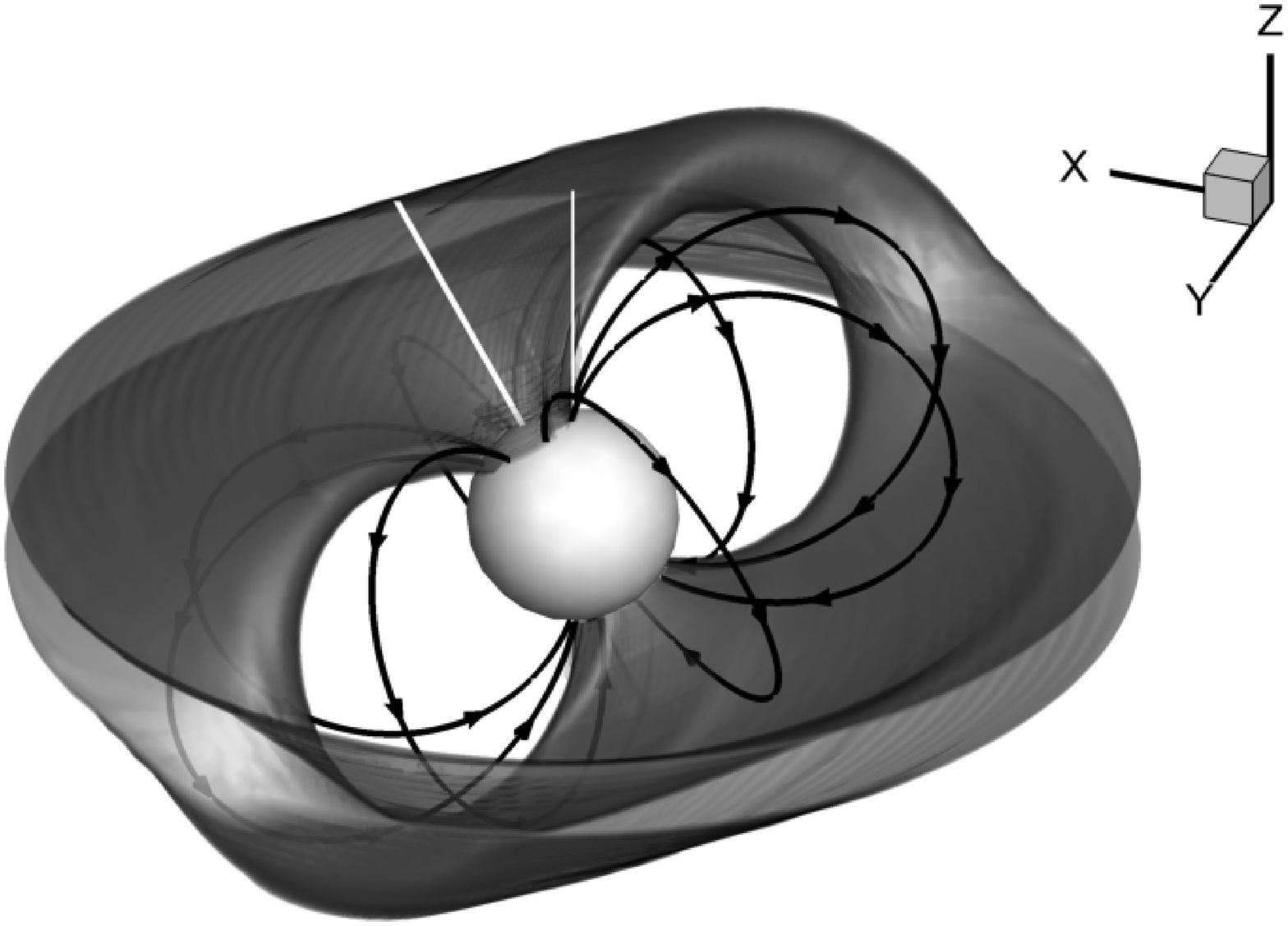}} \\
\end{tabular}
\caption{Flow structure (side view) for $B_a = 8~\text{kG}$ (left) and 
$B_a = 80~\text{kG}$ (right). Surfaces of constant logarithm of the density
are shown in the shade of gray, the magnetic field lines by lines with arrows, 
the rotation axis by the thin white solid line, and the
magnetic axis by the bold white solid line.}
\label{fig1}
\end{figure}

\begin{figure}[ht!]  
\centering 
\begin{tabular}{cc}
\hbox{\includegraphics[width=0.45\textwidth]{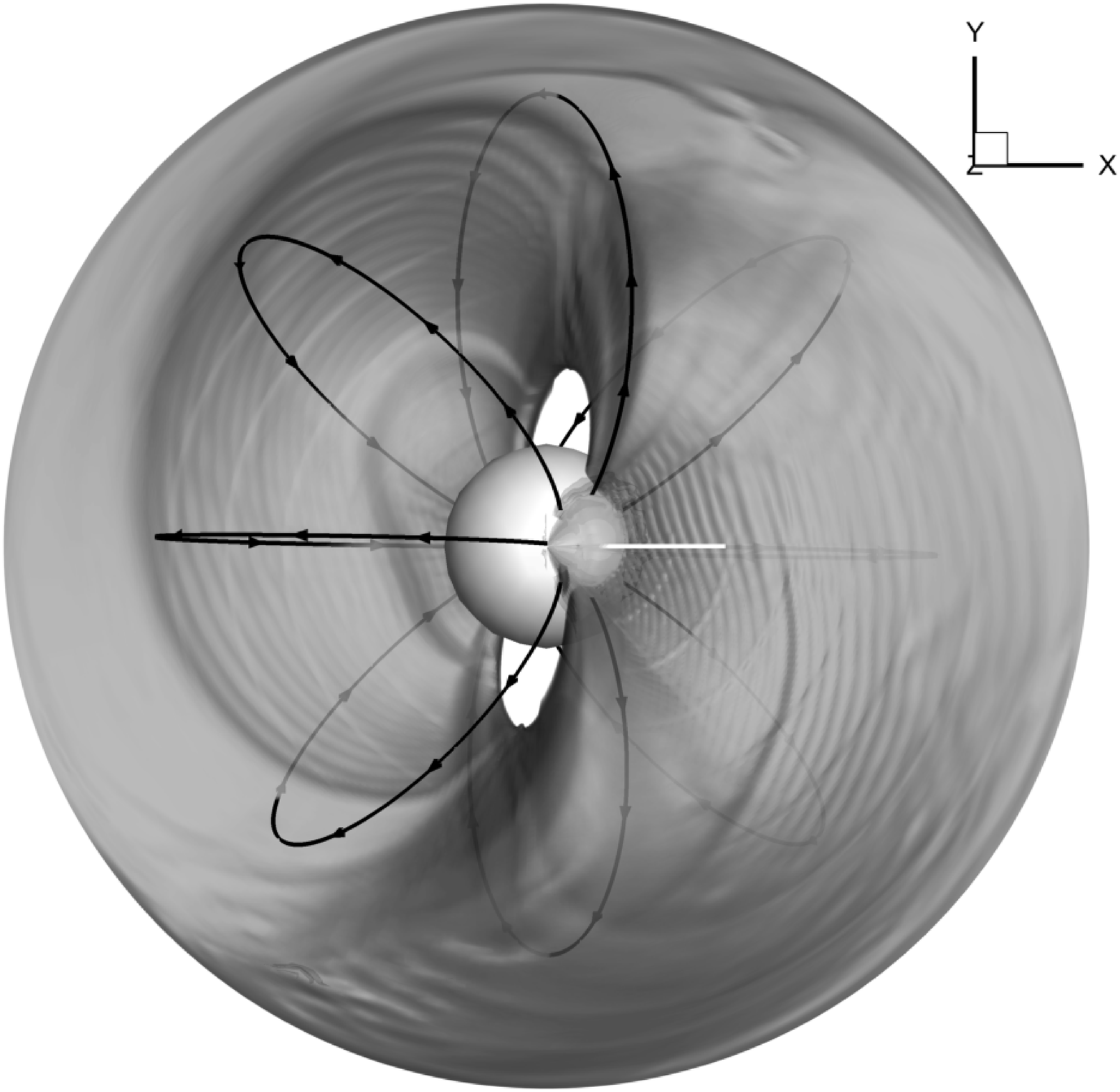}} &
\hbox{\includegraphics[width=0.45\textwidth]{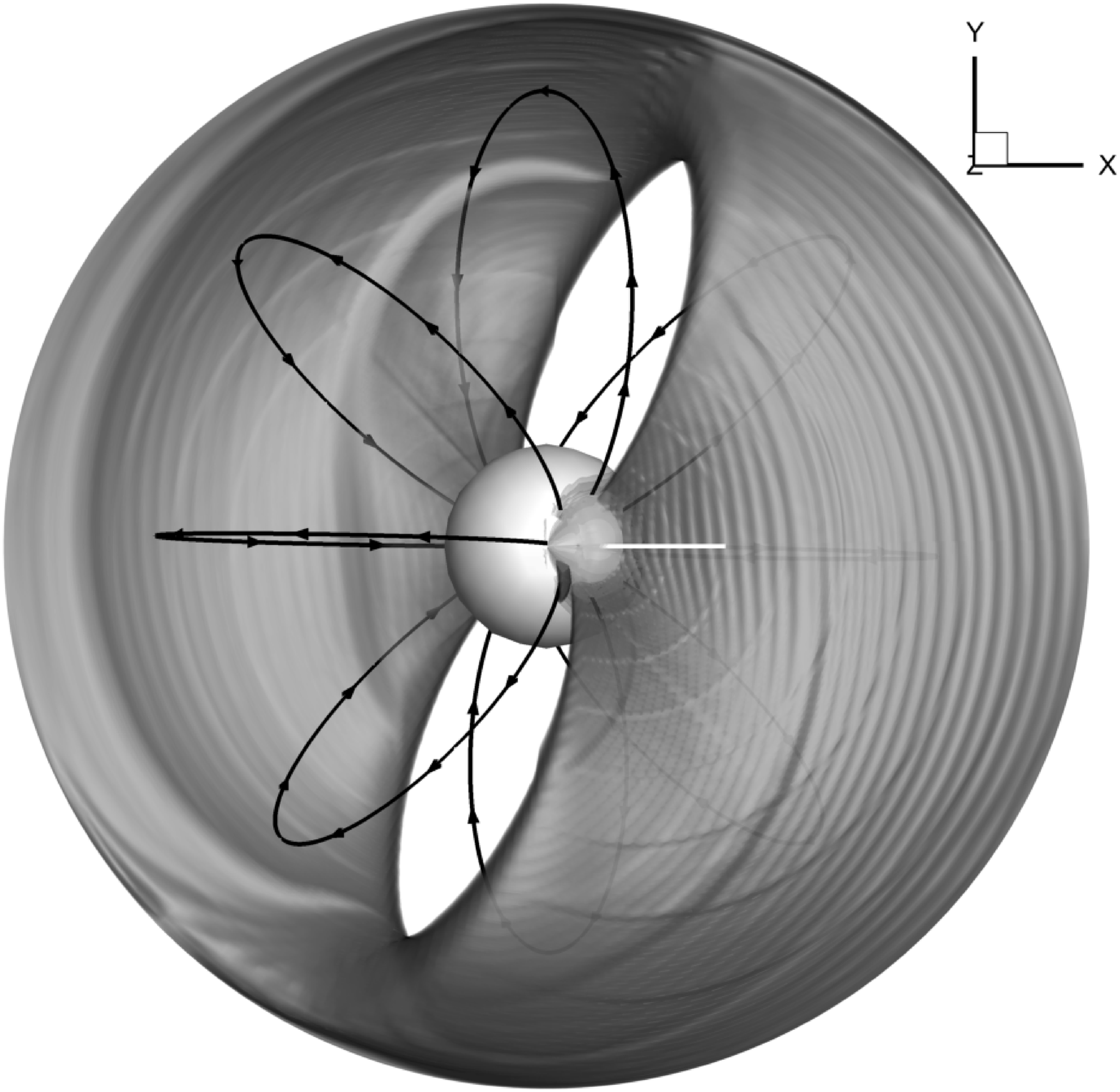}} \\
\end{tabular}
\caption{Same as Fig. \ref{fig1} shown from above.}
\label{fig2}
\end{figure}

The 3D structure of the flow is shown in Figs. \ref{fig1}
and \ref{fig2}. The left panels correspond to the Model 1, and the
right panels~--- to the Model 2. The bright sphere corresponds
to the accretor and the shade of gray shows isosurfaces
of logarithm of the density. The curved lines
with arrows indicate the direction of the magnetic 
field lines. The thin, white solid line is directed along
the $z$ axis and corresponds to the rotational axis of
the accretor, while the bold, white solid line shows the
magnetic axis.

Analysis of these figures shows the formation of a
magnetosphere near the surface of the accretor, where 
the matter moves mainly along magnetic field
lines. This results in column accretion, with the
matter reaching the surface of the accretor in the
vicinity of its magnetic poles. The accretion disk has
a nonuniform vertical structure. The disk thickness
decreases in places where the accretion columns begin
to form. Two cavities (vacuum regions) that are free
of matter form between the accretion disk and the
accretor; the magnetic field hinders the penetration
of matter into these regions, since the field lines pass
mainly along the stellar surface close to the
magnetic equator. The size of this vacuum region
increases with the field strength. Figure \ref{fig2} shows that
these regions are tilted by some angle; this is due to
the rotation of the matter in the disk, which causes
the accretion column shifts in the direction opposite
to the direction of rotation.

Figures \ref{fig1} and \ref{fig2} show that, in both cases, the 
accretion column has a curtain-like, rather than tube-
like, shape. The curtain is broader and more dense
in the Model 1, and its opening angle is almost equal to
$180^\circ$. The curtain occupies a much smaller volume
in the Model 2, and is narrower and less dense. In both
cases, the matter arrives to the surface of the white
dwarf in the shape of two arcs, forming hot spots where energy
is released.

\begin{figure}[ht!]  
\centering 
\begin{tabular}{cc}
\hbox{\includegraphics[width=0.45\textwidth]{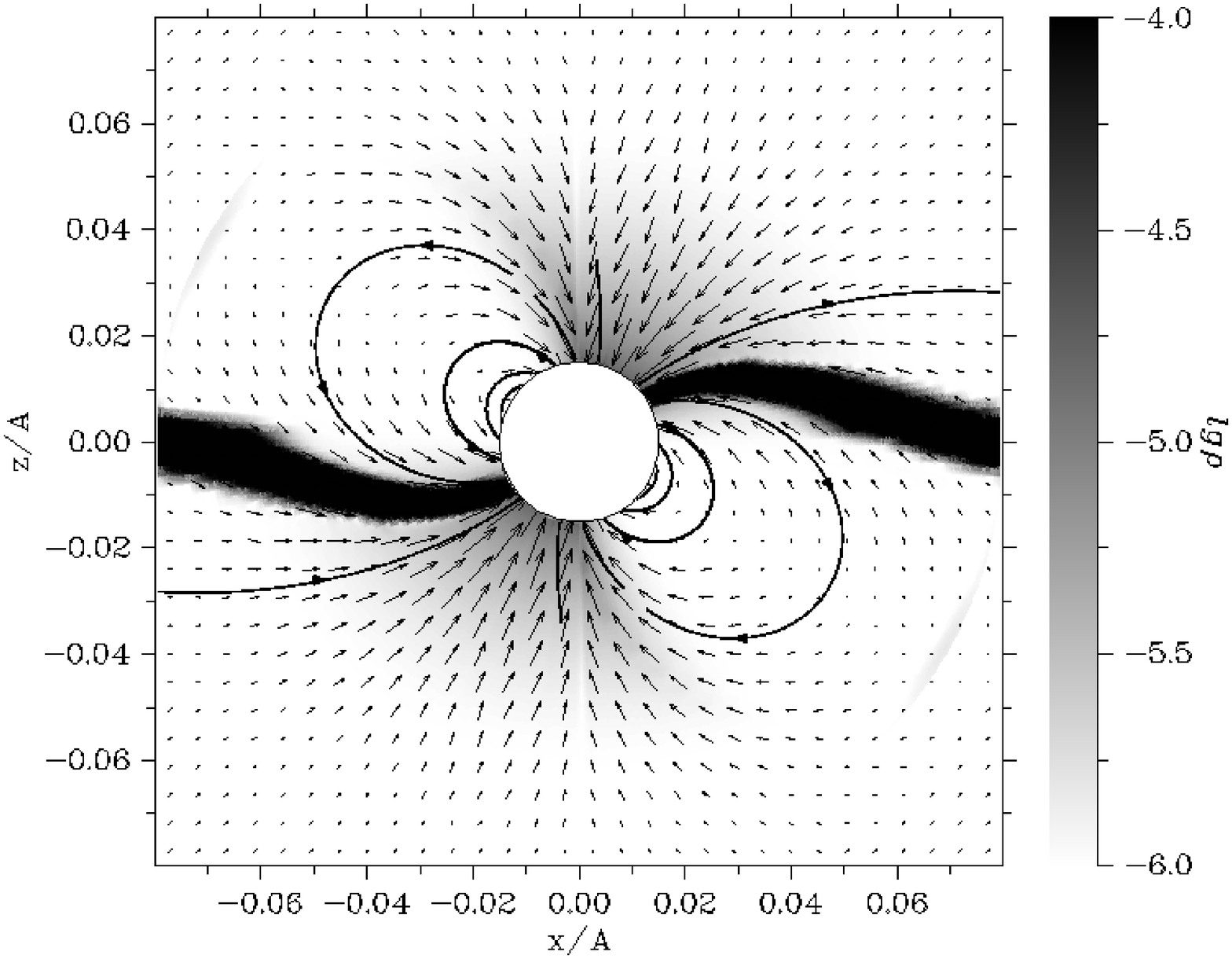}} &
\hbox{\includegraphics[width=0.45\textwidth]{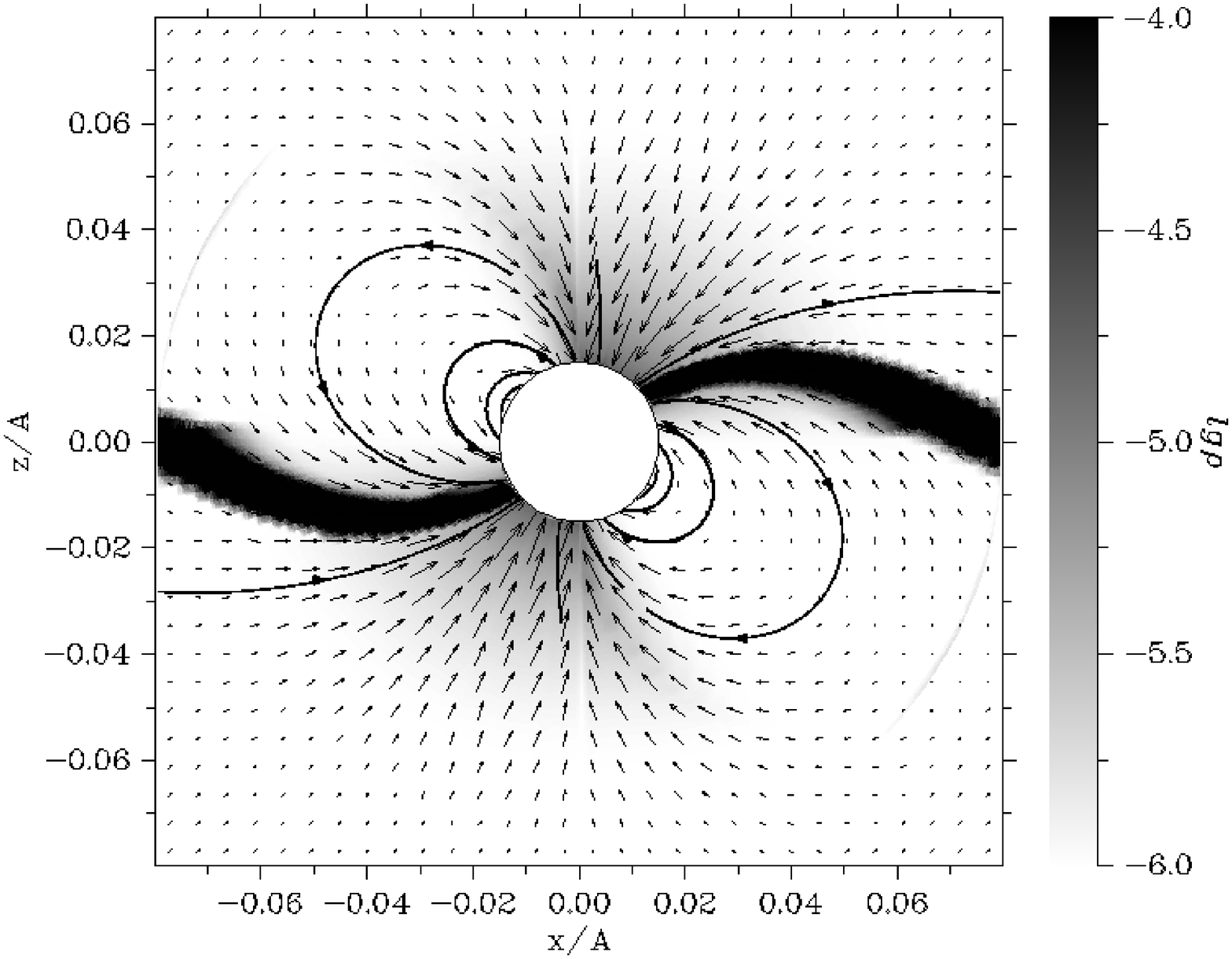}} \\
\end{tabular}
\caption{Flow structure in the vertical ($xz$) plane for the Model 1 (left) and Model 2 (right). 
The distribution of the logarithm of the
density is shown in the shade of gray in units of $\rho(\textrm L_1)$) together with the distribution of the 
velocity (arrows). The lines with arrows correspond to the magnetic field lines.}
\label{fig3}
\end{figure}

The flow structure in the vertical ($xz$) plane is
shown in Fig. \ref{fig3}. The left diagram corresponds to the
Model 1 and the right diagram~--- to the Model 2. The
distribution of the logarithm of the density (in units
of $\rho(\textrm L_1)$) is shown in the shade of gray. The arrows show
the velocity distribution, and the lines with arrows
show magnetic field lines. The flow pattern shown
is consistent with what was said above, and all features of the
flow structures noted above are clearly visible: the
magnetospheric region, accretion columns, and vacuum 
cavities. The difference in the flow structures for
these two models in the $xz$ plane is fairly weak. The
difference is manifest more clearly in an analysis of the
3D distributions.

\begin{figure}[ht]  
\centering 
\begin{tabular}{cc}
\hbox{\includegraphics[width=0.35\textwidth]{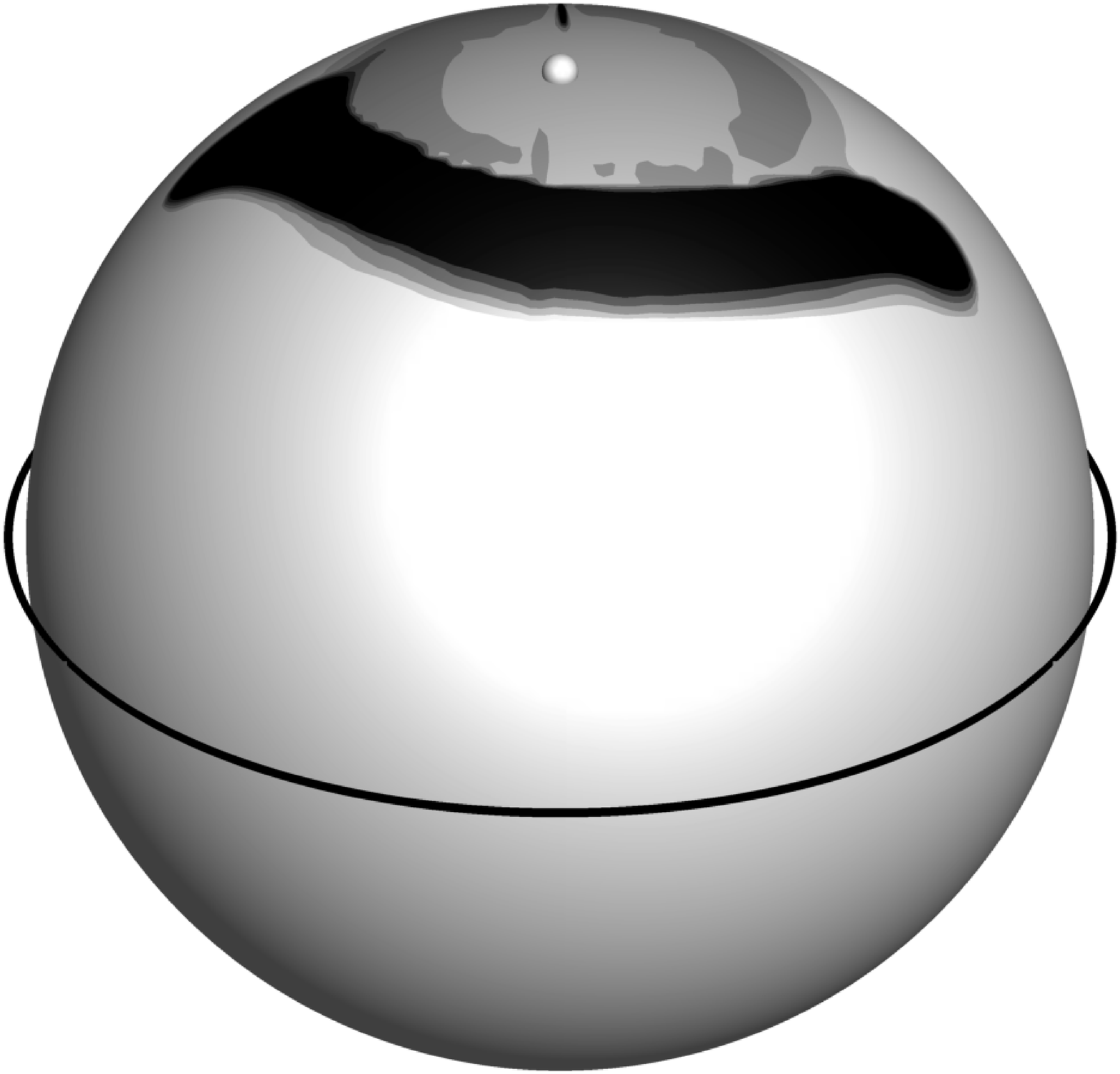}} &
\hbox{\includegraphics[width=0.35\textwidth]{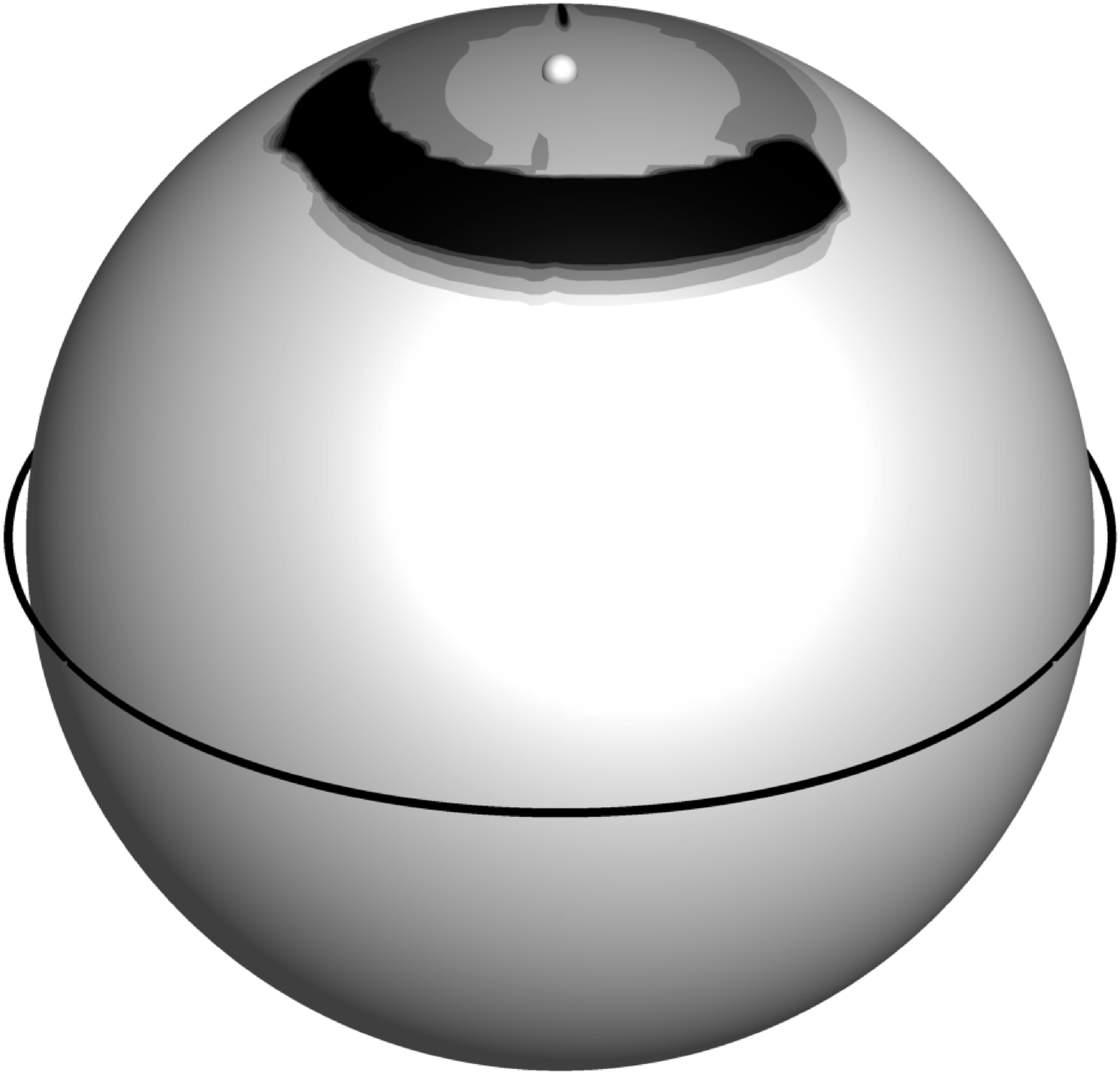}} \\
\hbox{\includegraphics[width=0.35\textwidth]{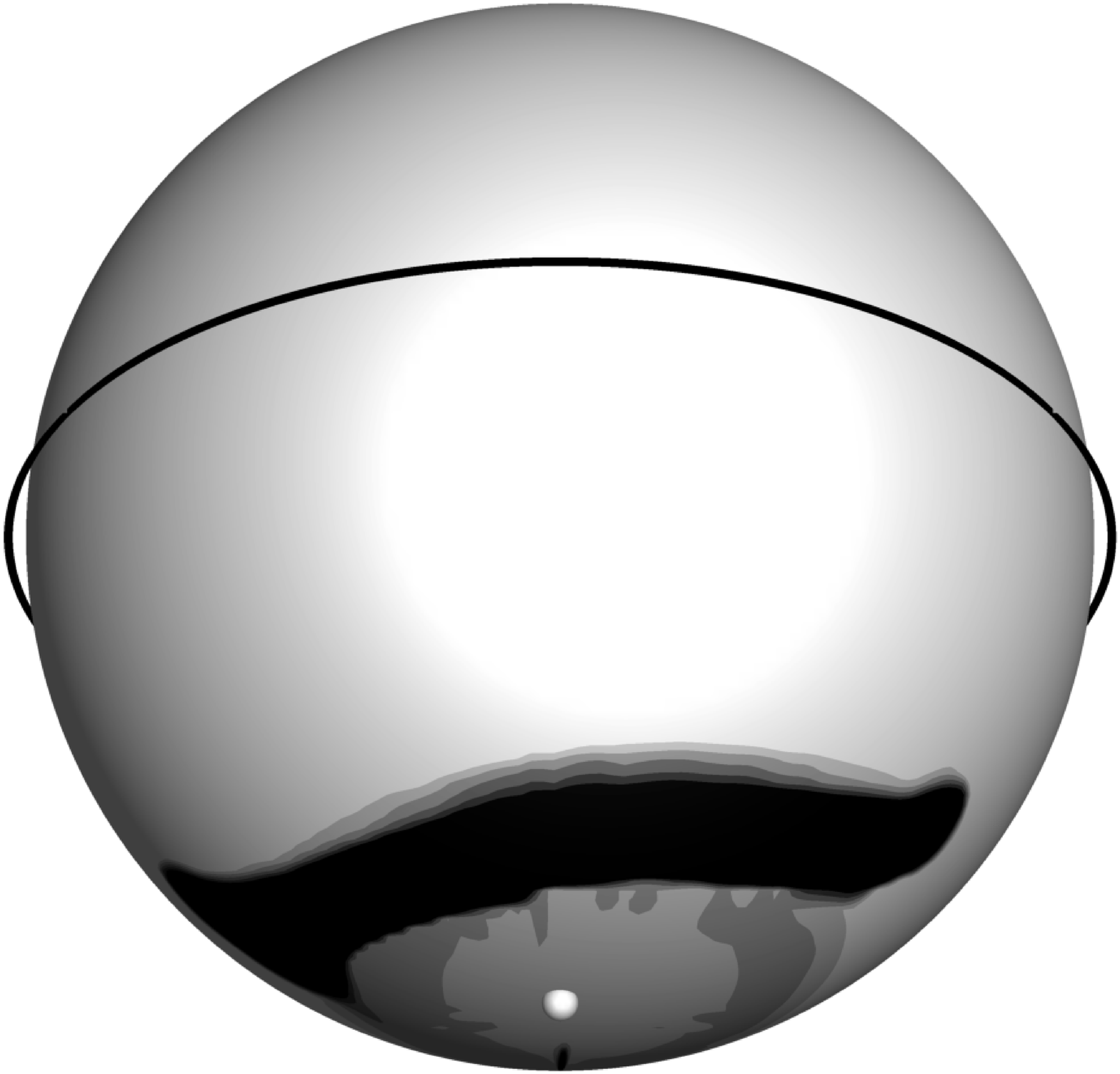}} &
\hbox{\includegraphics[width=0.35\textwidth]{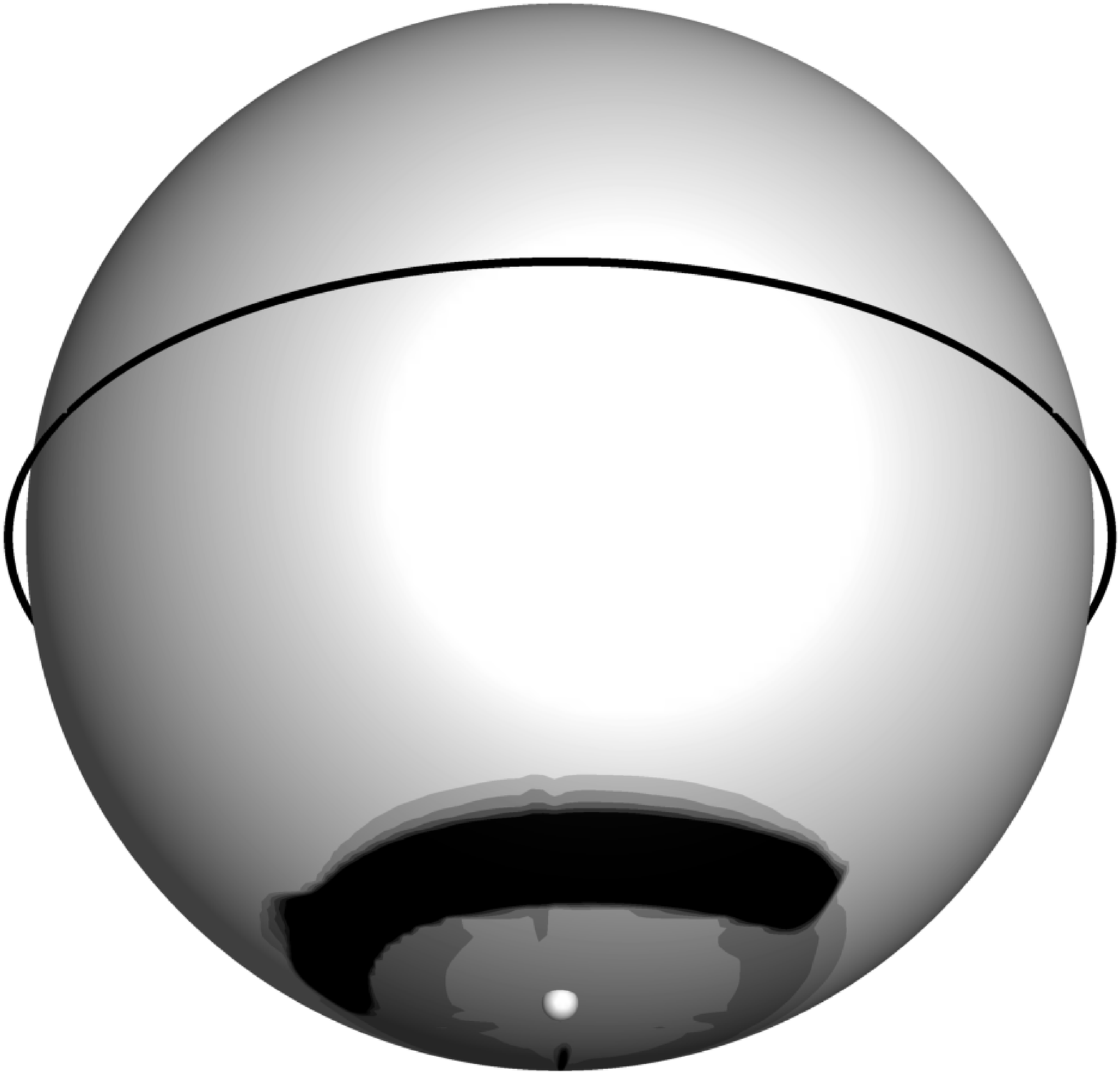}} \\
\end{tabular}
\caption{Distribution of the logarithm of the density at the surface of the 
accretor for $B_a = 8~\text{kG}$ (left) and $B_a = 8~\text{kG}$ (right).
The black line corresponding to the magnetic equator is shown. The positions of the 
Northern (top) and Southern (bottom) magnetic
poles are shown by the small white circles.}
\label{fig4}
\end{figure}

The shape of the hot spots is demonstrated in
Fig. \ref{fig4}, that presents the distributions of the 
logarithm of the density at the stellar surface for the Model 1
(left) and Model 2 (right). The upper diagrams focus
on the Northern hemisphere and hot spot, and the
lower diagrams on the Southern hemispheres and hot
spot. The white circles mark the positions of
the Northern and Southern magnetic poles (upper
and lower diagrams, respectively). The curved line
corresponds to the magnetic equator.

These figures show that the areas of energy release
have an arc-like shape (parts of an ellipse), that
is clearly due to the effect of gravity. For particles
moving along the magnetic field lines, it is energetically 
more profitable to fall onto the surface of the
accretor closer to the equator.
Therefore, the highest density is observed in precisely
these places. The falling of matter at the opposite 
end of the circumpolar accretion rings requires a
larger expense of energy.

The hot spot is more uniform and occupies a larger
area in the Model 1 (weaker magnetic field). This is
explained by the dependence of the wave magnetic
viscosity and the decay time on the magnetic field
strength $\eta_w \propto B_*$ in Eq. \eqref{eq-tw} and 
$t_w \propto 1/B_*$ and \eqref{eq-etaw} in
Eq. \eqref{eq-tw}. As a result, the force from the external magnetic
field acting on the plasma (the last term in Eq. (5))
is proportional to the field $B_*$. Therefore, the plasma
can more easily move across the magnetic field lines
in a weaker field (Model 1), and the accretion hot spot
spreads over a larger area.

Each spot in the Model 1 occupies about 7\% of the
stellar surface. The opening angles of the Northern 
and Southern spots are approximately $170^\circ$. In the
Model 2 (stronger magnetic field), the spot area is
smaller and the density distribution in the spot is more
nonuniform. Most of the accretion flow is concentrated 
toward the center of the spot. In Model 2, the
spot occupies about 4\% of the stellar surface area, and
the opening angle of the spots is about $140^\circ$.

\section{CONCLUSION}
We have developed a three-dimensional numerical
model that allows the detail studies of the flow structure 
near the surface of the accretor in a magnetic
close binary system. The model assumes that the
intrinsic magnetic field of the accretor is dipole, with
the dipole axis inclined to the rotational axis. The
model is based on the equations of modified 
magnetogasdynamics, that describe the mean 
characteristics of the flow in the frame of the wave MHD turbulence. This
approach performed well in our earlier calculations of
the flow structure in intermediate polars and polars.
The numerical model takes into account diffusion of
the magnetic field and radiative heating and cooling
processes.

We have presented here the results of 3D numerical 
simulations of accretion in a typical intermediate 
polar. The calculations were performed for
two intrinsic accretor magnetic fields~--- 8 kG and
80~kG~--- and an inclination of the magnetic axis
to the rotational axis of $30^\circ$. The results show the
formation of a magnetosphere close to the accretor,
with the accretion occurring through columns. The
accretion columns have a curtain-like rather than
tubular shape. The flow structure depends substantially 
on the field strength, although the picture does
not change qualitatively for different field strengths.
With increasing magnetic field strength, the 
magnetosphere expands, the vacuum regions become larger,
and the opening angles of the curtains decrease. The
zones of energy release (hot spots) at the surface
of the white dwarf that form in the vicinity of the
magnetic poles as a result of the matter inflow and they have
the shape of arcs or sections of an ellipse. Increasing
the field strength results in an increase in the hot spots area 
and a decrease in the their opening angles.

\medskip
This work was supported by the Russian Foundation for Basic Research (projects 14-29-06059,
14-02-00215, 15-02-06365), Basic Research Program P-41 of the Presidium of The Russian Academy
of Sciences, by the President of the Russian Federation Grant NSh-
3620.2014.2).

\end{document}